\documentclass[11pt]{article}

\usepackage{a4}
\usepackage{bbm}

\usepackage{amssymb} 
\usepackage{amsmath} 
\usepackage{amsthm} 
\usepackage{setspace} 
\usepackage{cite} 
\usepackage{hyperref}
\usepackage{verbatim}
\usepackage{graphicx}
\usepackage{subfig}
\usepackage{fancybox}
\usepackage{microtype}
\usepackage{latexsym}

\usepackage{tikz}

\usetikzlibrary{calc,positioning,shapes,shadows,arrows,fit}

\usetikzlibrary{arrows.meta}

\DeclareFontEncoding{LS1}{}{}
\DeclareFontSubstitution{LS1}{stix}{m}{n}
\DeclareSymbolFont{symbols2stix}      {LS1}{stixfrak} {m} {n}
\DeclareMathSymbol{\lparenless}               {\mathopen} {symbols2stix}{"32}
\DeclareMathSymbol{\rparengtr}                {\mathclose}{symbols2stix}{"33}

\newcommand{\newbrak}[1]{{\lparenless} #1 {\rparengtr}}
\newtheorem{definition}{Definition} 
\newtheorem{corollary}{Corollary} 
 
\newtheorem{claim}{Claim} 
\newtheorem{lemma}{Lemma} 
\newtheorem{theorem}{Theorem} 
 
\newtheorem{obs}{Observation}

\newtheorem{proposition}{Proposition}
\newtheorem{remark}{Remark}

\usepackage{thm-restate}
\usepackage{hyperref}

\usepackage{cleveref}

\newcommand{\F}{\mathbb{F}}

\renewcommand{\angle}[1]{{\langle} #1 {\rangle}}

\newcommand{\cproof}{\noindent{\it Proof of Claim}}

\newcommand{\gX}{X,X^{-1}} 
\newcommand{\ratX}{\mathbb{F}\newbrak{X}}

\newcommand{\Ex}{\mathcal{E}}

\DeclareMathOperator{\poly}{\mbox{\small\rm poly}}

\title{Efficient Black-Box Identity Testing for Free Group Algebra}

\author{V. Arvind\thanks{Institute of Mathematical Sciences (HBNI), Chennai,
India, \texttt{email: arvind@imsc.res.in}}  \and Abhranil Chatterjee\thanks{Institute of Mathematical Sciences (HBNI), Chennai,
    India, \texttt{email: abhranilc@imsc.res.in}} \and Rajit Datta\thanks{Chennai Mathematical Institute, Chennai, India, \texttt{email: rajit@cmi.ac.in}} \and Partha
  Mukhopadhyay\thanks{Chennai Mathematical Institute, Chennai, India,
    \texttt{email: partham@cmi.ac.in}}
}

\begin{document} 
\maketitle

\begin{abstract}
  Hrube\v s and Wigderson \cite{HW14} initiated the study of
  noncommutative arithmetic circuits with division computing a
  noncommutative rational function in the \emph{free skew field}, and
  raised the question of rational identity testing. It is now known
  that the problem can be solved in deterministic polynomial time in
  the \emph{white-box} model for noncommutative formulas with
  inverses, and in randomized polynomial time in the \emph{black-box}
  model \cite{GGOW16, IQS18, DM18}, where the running time is
  polynomial in the size of the formula. 

  The complexity of identity testing of noncommutative rational
  functions remains open in general (when the formula size
  is not polynomially bounded). We solve the problem for a natural
  special case. We consider polynomial expressions in the free group
  algebra $\F\angle{\gX}$~\footnote{We use $\F\angle{X,X^{-1}}$ to denote
  $\F\angle{x_1,\ldots,x_n,x^{-1}_1,\ldots,x^{-1}_n}$.} where $X=\{x_1, x_2, \ldots, x_n\}$, a
  subclass of rational expressions of inversion height one. Our main
  results are the following.
\begin{enumerate}
\item Given a degree $d$ expression $f$ in $\F\angle{\gX}$ as a
  black-box, we obtain a randomized $\poly(n,d)$ algorithm to check
  whether $f$ is an identically zero expression or not. We obtain this
  by generalizing the Amitsur-Levitzki theorem \cite{AL50} to
  $\F\angle{\gX}$. This also yields a deterministic identity testing algorithm (and even
  an expression reconstruction algorithm) that is polynomial time in
  the sparsity of the input expression.
\item Given an expression $f$ in $\F\angle{\gX}$ of degree at most
  $D$, and sparsity $s$, as black-box, we can check whether $f$ is
  identically zero or not in randomized $\poly(n,\log s, \log D)$
  time.
\end{enumerate}   
\end{abstract}

\section{Introduction}

Noncommutative computation is an important sub-area of arithmetic
circuit complexity. In the usual arithmetic circuit model for
noncommutative computation, the arithmetic operations are addition and
multiplication. However, the multiplication gates respect the input
order since the variables are noncommuting. Analogous to commutative
arithmetic computation, the central questions are to show lower bounds
for explicit polynomials and derandomization of polynomial identity
testing (PIT) for noncommutative polynomial rings. Exploiting the
limited cancellations, strong lower bounds and PIT results are known
for noncommutative computations (in contrast to the commutative
setting). Nisan\cite{Ni91} has shown that any algebraic branching
program (ABP) computing the $n\times n$ noncommutative Determinant or
Permanent polynomial requires exponential (in $n$) size. On the PIT
front, Raz and Shpilka~\cite{RS05} have shown a deterministic
polynomial-time PIT for noncommutative ABPs in the white-box model. A
quasi-polynomial time derandomization is also known for the black-box
model~\cite{FS12}. However, for general circuits there are no better
results (either lower bound or PIT) than known in the commutative
setting.

The randomized polynomial-time PIT algorithm for noncommutative
circuits computing a polynomial of polynomially bounded
degree~\cite{BW05} follows from Amitsur-Levitzki theorem~\cite{AL50}.
The Amitsur-Levitzki theorem states that a nonzero noncommutative
polynomial $p\in\F\angle{X}$ of degree $<2k$ cannot be an identity for
the matrix ring $\mathbb{M}_k(\F)$.  Additionally, it is shown that a
nonzero noncommutative polynomial does not vanish on matrices of
dimension logarithmic in the sparsity of the polynomial, yielding a
randomized polynomial time algorithm for noncommutative circuits
computing a nonzero polynomial of exponential degree and exponential
sparsity~\cite{AJMR17}.

Hrube\v s and Wigderson \cite{HW14} initiated the study of
noncommutative computation with inverses.  In the commutative world,
it suffices to consider additions and multiplications. By Strassen's
result \cite{Str73} (extended to finite fields \cite{HY11a}),
divisions can be efficiently replaced by polynomially many additions
and multiplications. However, divisions in noncommutative computation
are more complex~\cite{HW14}. In the same paper~\cite{HW14} the
authors introduce \emph{rational identity testing}: Given a
noncommutative formula involving addition, multiplication and
division gates, efficiently check if the resulting rational expression
is identically zero in the free skew-field of noncommutative rational
functions. They show that the rational identity testing problem
reduces to the following SINGULAR problem:

Given a matrix $A_{n \times n}$ where the entries are linear forms
over noncommuting variables $\{ x_1,x_2,\ldots,x_n \}$, is $A$
invertible in the free skew-field?

In the white-box model the problem is in deterministic polynomial
time, and in randomized polynomial time in the black-box model
~\cite{GGOW16,IQS18,DM18}. Specifically, for rational formulas of size
$s$, random matrix substitutions of dimension linear in $s$ suffices
to test if the rational expression is identically zero \cite{DM18}.

The complexity of identity testing for general rational expressions
remains open. For example, given a
noncommutative circuit involving addition, multiplication and
division gates, no efficient algorithm is known to check if the resulting rational expression
is identically zero in the free skew-field of noncommutative rational
functions. In order to precisely formulate the problem, we define
classes of rational expressions based on Bergman's definition
\cite{Ber76} of \emph{inversion height} which we now recall and
elaborate upon with some notation.

\begin{definition}{\rm\cite{Ber76}}
  Let $X$ be a set of free noncommuting variables. Polynomials in the
  free ring $\F\angle{X}$ are defined to be rational expressions of
  \emph{height $0$}. A rational expression of \emph{height $i+1$} is
  inductively defined to be a polynomial in rational expressions of
  height at most $i$, and inverses of such expressions.
\end{definition}

Let $\Ex_{d,0}$ denote all polynomials of degree at most $d$ in the
free ring $\F\angle{X}$. We inductively define rational expressions in
$\Ex_{d,i+1}$ as follows: Let $f_1,f_2,\ldots,f_r$ and
$g_1,g_2,\ldots,g_s$ be rational expressions in $\Ex_{d,i}$ in the
variables $x_1,x_2,\ldots,x_n$. Let
$f(y_1,y_2,\ldots,y_s,z_1,z_2,\ldots,z_r)$ be a degree-$d$ polynomial
in $\F\angle{X}$. Then
$f(g_1,g_2,\ldots,g_s,f^{-1}_1,f^{-1}_2,\ldots,f^{-1}_r)$ is a
rational expression (of inversion height $i+1$) in $\Ex_{d,i+1}$.

Black-box identity testing for rational expressions is not well
understood in general. Bergman has shown \cite[Proposition 5.1]{Ber76}
that there are rational expressions that are nonzero over a dense
subset of $2\times 2$ matrices but evaluate to zero on dense subsets
of $3\times 3$ matrices. This makes it difficult to formulate an
Amitsur-Levitzki type of theorem\cite{AL50} for rational expressions.

\begin{remark}
  In this connection, we note that Hrube\v s and Wigderson \cite{HW14}
  have observed that testing if a `correct' rational expression $\Phi$ is not
  identically zero is equivalent to testing if the rational expression
  $\Phi^{-1}$ is `correct'. I.e. testing if a correct rational
  expression of \emph{inversion height $i$} is identically zero or not can be reduced
  to testing if a rational
  expression of \emph{inversion height $i+1$} is correct or not.
  Furthermore, testing if a rational
  expression of \emph{inversion height one} is correct can be done by
  applying (to each inversion operation in this expression) a theorem
  of Amitsur (see~\cite{row80, LZ09}) which implies that a nonzero degree $2d-1$ noncommutative
  polynomial evaluated on $d\times d$ matrices will be invertible with
  high probability. However, this does not yield an efficient
  randomized identity testing algorithm for rational expressions of
  inversion height one. Because that seems to require testing
  correctness of expressions of inversion height two which is a
  question left open in their paper \cite[Section 9]{HW14}.
\end{remark}

\subsubsection*{The Free Group Algebra}

This motivates the study of black-box identity testing for rational
expressions in the \emph{free group algebra} $\F\angle{\gX}$. 

We consider expressions in the free group algebra $\F\angle{\gX}$,
where $(\gX)^*$ denotes the free group generated by the $n$ generators
$X=\{x_1,x_2,\ldots,x_n\}$ and their inverses
\[
X^{-1} =\{x_1^{-1},x_2^{-1},\ldots,x_n^{-1}\}.
\]
Elements of the free group $(\gX)^*$ are words in $\gX$. The only
relations satisfied by the generators is $x_ix_i^{-1}=x_i^{-1}x_i=1$
for all $i$. Thus, the elements in the free group $(\gX)^*$ are the
\emph{reduced words} which are words to which the above relations are
not applicable.

The elements of the \emph{free group algebra} $\F\angle{\gX}$ are
$\F$-linear combinations of the form 
\[
f = \sum_w \alpha_w w,~~\alpha_w\in\F,
\]
where each $w\in (\gX)^*$ is a reduced word. The \emph{degree} of the
expression $f$ is defined as the maximum length of a word $w$ such
that $\alpha_w\ne 0$. The expression $f$ is said to have
\emph{sparsity} $s$ if there are $s$ many reduced words $w$ such that
$\alpha_w\ne 0$ in $f$. We also use the notation $[w]f$ to denote the
coefficient $\alpha_w$ of the reduced word $w$ in the expression $f$.

The free noncommutative ring $\F\angle{X}$ is a subalgebra of
$\F\angle{\gX}$. Clearly, the elements of $\F\angle{\gX}$ are a
special case of rational expressions of \emph{inversion height
  one}. I.e., we note that:

\begin{proposition}
$\F\angle{\gX}\subset \cup_{d>0}\Ex_{d,1}$.
\end{proposition} 

Note that the rational expressions in $\F\angle{\gX}$ allows inverses
only of the variables $x_i$, whereas the \emph{free skew field}
$\ratX$ contains all possible rational expressions (with inverses at
any nested level).

\subsection*{Our results}
The main goal of the current paper is to
obtain black-box identity tests for rational expressions in
the free group algebra $\F\angle{\gX}$.

Our first result is a generalization of the Amitsur-Levitzki
theorem\cite{AL50} to $\F\angle{X,X^{-1}}$. Let $A$ be an associative
algebra with identity over $\F$. An expression $f \in
\F\angle{X,X^{-1}}$ is an
\emph{identity} for $A$ if
\[
f(a_1,\ldots,a_n)= 0 
\]
for all $a_i \in A$ such that $a^{-1}_i$ is defined for each $i\in[n]$.

\begin{theorem}\label{PI-theorem}
Let $\F$ be any field of characteristic zero and $f\in\F\angle{\gX}$ be a nonzero expression of degree $d$. 
Then $f$ is not an identity
for the matrix algebra $\mathbb{M}_{2d}(\F)$.
\end{theorem} 

The following corollary is immediate.
 
\begin{corollary}[Black-box identity testing for circuits in free group algebra]\label{rand-nccirc}
There is a black-box randomized $\poly(n,d)$ identity test for degree
$d$ expressions in $\F\angle{X,X^{-1}}$.
\end{corollary}

If the black-box contains a sparse expression, we show efficient
deterministic algorithms for identity testing and interpolation
algorithm.
 
\begin{theorem}[Black-box identity testing and reconstruction for sparse expressions in free group algebra]\label{det-sparsenc}
Let $\F$ be any field of characteristic zero and $f$ is an expression in $\F\angle{X,X^{-1}}$ of degree $d$ and
sparsity $s$ given as black-box. Then we can reconstruct $f$ in deterministic
$\poly(n,d,s)$ time with matrix-valued queries to the black-box.
\end{theorem}

Our next result is another generalization of the Amitsur-Levitzki
theorem~\cite{AL50} extending a result of \cite{AJMR17} to free group
algebras. We show that a nonzero expression $f\in\F\angle{\gX}$ of
degree $D$ and sparsity $s$ does not vanish on $O(\log s)$ dimensional
matrices. It yields a randomized polynomial-time identity test if the
black-box contains an expression $f$ of exponential degree and
exponential sparsity.

\begin{theorem}\label{sparse-PI-theorem}
Let $\F$ be any field of characteristic zero. 
Then, a degree-$D$ expression $f\in\F\angle{\gX}$ of sparsity $s$ is not an identity for the matrix algebra $\mathbb{M}_k(\F)$ for $k = O(\log s)$.
\end{theorem}

\begin{corollary}[Black-box identity testing for expoential sparse expressions with exponential degree in free group algebra]\label{expo-deg-sparse-pit-theorem}
Given a degree-$D$ expression $f\in \F\angle{X,X^{-1}}$ of sparsity
$s$ as black-box, we can check whether $f$ is identically zero or not
in randomized $\poly(n,\log D, \log s)$ time.
\end{corollary}


\begin{remark}\label{finite-field-case}
We state our results for fields of characteristic zero only for simplicity.
However, by suitable modifications, we can extend our results for fields of positive characteristic.
\end{remark}

\subsection*{Organization}
The paper is organized as follows. In Section \ref{main-sect}, we
prove Theorem \ref{PI-theorem}, Corollary \ref{rand-nccirc}, and
Theorem \ref{det-sparsenc}. In Section \ref{fga-expodeg-pit}, we prove
Theorem \ref{sparse-PI-theorem} and Corollary
\ref{expo-deg-sparse-pit-theorem}. Finally, in Section \ref{finite-field-modifications}, we discuss suitable modifications to extend our results over finite fields.


\section{A Generalization of Amitsur-Levitzki Theorem for Free Group Algebra}\label{main-sect}

The main idea in our proof is to efficiently encode expressions in
$\F\angle{X,X^{-1}}$ as polynomials in a suitable commutative ring
preserving the identity. Let $\F[Y, Z]$ denote the commutative ring
$\F[y_{ij},z_{ij}]_{i\in [n],j\in [d]}$ for $n,d\in \mathbb{N}$,
where $Y=\{y_{ij}\mid i\in[n], j\in[d]\}$ and $Z=\{z_{ij}\mid i\in[n],
j\in[d]\}$.

\begin{definition}\label{defn-encoding}
Define a map $\varphi : \F\angle{X,X^{-1}}\to \F[Y,Z]$ to be a map
such that $\varphi$ is identity on $\F$, and for each reduced
word $w=x^{b_1}_{i_1}x^{b_2}_{i_2}\cdots x^{b_d}_{i_d}$,
\[
\varphi(x^{b_1}_{i_1}x^{b_2}_{i_2}\cdots x^{b_d}_{i_d}) = \prod_{j=1}^d (\mathbbm{1}_{[b_j=1]}\cdot y_{i_jj} + \mathbbm{1}_{[b_j=-1]} \cdot z_{i_jj} ),
\]
where $\mathbbm{1}_{[b_j=b]}=1$ if $b_j=b$ and
$\mathbbm{1}_{[b_j=b]}=0$ otherwise.
\end{definition}

By linearity the map $\varphi$ is defined on all expressions in
$\F\angle{\gX}$. We observe the following properties of $\varphi$.
  
\begin{enumerate}
\item The map $\varphi$ is injective on the reduced words $(\gX)^*$.
  I.e., it maps each reduced word $w\in(\gX)^*$ to a unique monomial
  over the commuting variables $Y\cup Z$.
\item Consequently, $\varphi$ is identity preserving. I.e., an
  expression $f$ in $\F\angle{X,X^{-1}}$ is identically zero if and
  only if its image $\varphi(f)$ is the zero polynomial in $\F[Y,Z]$.
\item $\varphi$ preserves the sparsity of the expression. I.e., $f$ in
  $\F\angle{X,X^{-1}}$ is $s$-sparse iff $\varphi(f)$ in $\F[Y,Z]$ is
  $s$-sparse.
\item Given the image $\varphi(f)\in\F[Y,Z]$ in its sparse description
  (i.e., as a linear combination of monomials), we can efficiently
  recover the sparse description of $f\in\F\angle{\gX}$.
\end{enumerate}

Given polynomials $f,f'\in \F[Y,Z]$, we say $f$ and $f'$ are
\emph{weakly equivalent}, if for each monomial $m$, $[m]f = 0$ if and
only if $[m]f' = 0$, where $[m]f$ denotes the coefficient of monomial
$m$ in $f$.

Given a black-box expression $f$ in $\F\angle{X,X^{-1}}$, we show how
to evaluate it on suitable matrices and obtain a polynomial in
$\F[Y,Z]$ that is \emph{weakly equivalent to} $\varphi(f)$ as a
specific entry of the resulting matrix. The matrix substitutions are
based on automata constructions. Similar ideas have been used earlier
to design PIT algorithms for noncommutative polynomials \cite{AMS10}. However,
since we are dealing with rational expressions, some difficulties
arise. The matrix substitutions for the variables $x_1, \ldots, x_n$
are obtained as the corresponding transition matrices $M_i$ of the
automaton. The matrix substitution for $x_i^{-1}$ will be
$M_i^{-1}$. Therefore, we need to ensure that the transition matrices
$M_i$ are invertible and sufficiently structured to be useful for the
identity testing.

We first illustrate our construction for an example degree-$2$
expression $f=x_1x^{-1}_2 + x_2 x^{-1}_1$, where $X=\{x_1,x_2\}$.

The basic ``building block'' for the transition matrix $M_i$ is the
$2\times 2$ block matrix
\[
\begin{bmatrix}
0  &y_{ij}\\
\frac{1}{z_{ij}}  &0\\
\end{bmatrix},
\]
whose inverse is
\[
\begin{bmatrix}
0  &z_{ij}\\
\frac{1}{y_{ij}}  &0\\
\end{bmatrix}.
\]

When the $2\times 2$ block is the $j^{th}$ diagonal block in $M_i$,
the corresponding automaton will go from state $2j-1$ to state $2j$
replacing $x_i$ by $y_{ij}$ (or if $x_i^{-1}$ occurs, it will replace
it by $z_{ij}$).

We will keep the transition matrix $M_i$ for $x_i$ a block diagonal
matrix with such $2\times 2$ invertible blocks as the principal minors
along the diagonal. In order to ensure this we introduce two new
variables $W=\{w_1,w_2\}$ and substitute $x_i$ by the word $w_ix_iw_i$
in the expression. This will ensure that we do not have two
consecutive $x_i$ in the resulting reduced words. In fact, between two
$X$ variables (or their inverses) we will have inserted exactly two
$W$ variables (or their inverses). Now, we define $M_i$ for the
above example as

\[
M_i = 
\begin{bmatrix}

0  &y_{i1} &0 &0\\
\frac{1}{z_{i1}}  &0 &0 &0\\
0 &0 &0 &y_{i2}\\
0 &0 &\frac{1}{z_{i2}} &0\\
\end{bmatrix}, 
\quad\quad
M^{-1}_i = 
\begin{bmatrix}

0  &z_{i1} &0 &0\\
\frac{1}{y_{i1}}  &0 &0 &0\\
0 &0 &0 &z_{i2}\\
0 &0 &\frac{1}{y_{i2}} &0\\
\end{bmatrix}.
\] 

\begin{figure}[h]
\begin{center}
\begin{tikzpicture}
\node(pseudo) at (-1,0){};
\node(1) at (-3.5,0)[shape=circle,draw]        {$q_1$};
\node(2) at (.5,0)[shape=circle,draw]        {$q_2$};
\node(3) at (3.5,0)[shape=circle,draw] {$q_3$};
\node(4) at (7.5,0)[shape=circle,draw] {$q_4$};
\path [->]


  
  (1)      edge [bend right=-25]  node [above]  {$x_i\rightarrow y_{i1}$}     (2)
  (1)      edge [bend right=-65]  node [above]  {$x^{-1}_i\rightarrow  z_{i1}$}     (2)
  (2)      edge [bend left=25]  node [below]  {$x_i, x_i^{-1}\rightarrow 1/z_{i1}, 1/y_{i1}$}     (1)
  
  (2)      edge [dotted]  node [above]  {}     (3)
  
  (3)      edge [bend right=-25]  node [above]  {$x_i\rightarrow y_{i2}$}     (4)
  (3)      edge [bend right=-65]  node [above]  {$x^{-1}_i\rightarrow z_{i2}$}     (4)
  (4)      edge [bend left=25]  node [below]  {$x_i, x_i^{-1}\rightarrow 1/ z_{i2}, 1/y_{i2}$}     (3);
  


\end{tikzpicture}
\caption{The transition diagram of the automaton for $x$ variables}\label{fig3}
\end{center} 
\end{figure}
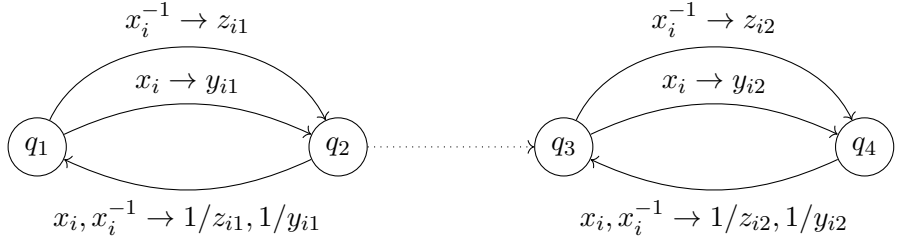

The corresponding transitions of the automaton is shown in
Figure~\ref{fig3}.

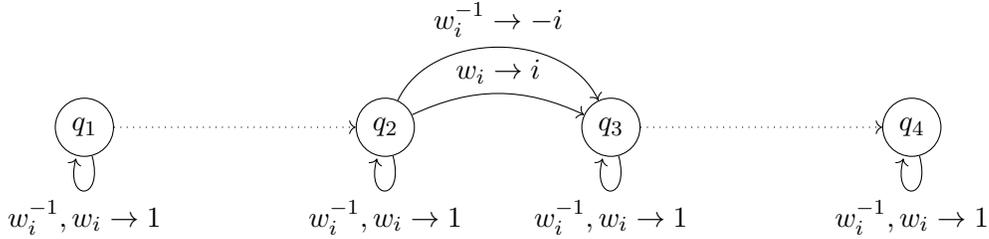
\begin{figure}[h]
\begin{center}
\begin{tikzpicture}
\node(pseudo) at (-1,0){};
\node(1) at (-3.5,0)[shape=circle,draw]        {$q_1$};
\node(2) at (.5,0)[shape=circle,draw]        {$q_2$};
\node(3) at (3.5,0)[shape=circle,draw] {$q_3$};
\node(4) at (7.5,0)[shape=circle,draw] {$q_4$};
\path [->]

(1)      edge [dotted]  node [above]  {}     (2)
  (3)      edge [dotted]  node [above]  {}     (4)
  (2)      edge [bend right=-25]  node [above]  {$w_i\rightarrow i$}     (3)
  (2)      edge [bend right=-65]  node [above]  {$w^{-1}_i\rightarrow -i$}     (3)
    
(1)      edge [loop below]    node [below]  {$w^{-1}_i,w_i \rightarrow 1$}     ()
(2)      edge [loop below]    node [below]  {$w^{-1}_i ,w_i\rightarrow 1$}     ()
(3)      edge [loop below]    node [below]  {$w^{-1}_i, w_i \rightarrow 1$}     ()
(4)      edge [loop below]    node [below]  {$w^{-1}_i , w_i\rightarrow 1$}     ();

\end{tikzpicture}
\caption{The transition diagram of the automaton for $w$ variables}\label{fig4}
\end{center} 
\end{figure}

We now describe the transition matrices $N_i$ for $w_i$. The matrix
$N_i$ is also a $4\times 4$ block diagonal matrix. There are three
blocks along the diagonal. The first and third are $1\times 1$ blocks
of the identity. The second one is a $2 \times 2$ block for
$w_i$-transitions from state $q_2$ to state $q_3$. It ensures that for
any subword $w^{b_1}_1 w^{b_2}_2$, $b_i\in \{1,-1\}$, in the resulting
product matrix $N^{b_1}_1 N^{b_2}_2$ the $(1,2)^{th}$ entry of the
$2\times 2$ block is nonzero. The corresponding transitions of the
automaton is depicted in Figure~\ref{fig4}.

\[
N_i = 
\begin{bmatrix}

1  &0 &0 &0\\
0  &1 &i &0\\
0 &0 &1 &0\\
0 &0 &0 &1\\
\end{bmatrix}
,\quad
N^{-1}_i = 
\begin{bmatrix}

1  &0 &0 &0\\
0  &1 &-i &0\\
0 &0 &1 &0\\
0 &0 &0 &1\\
\end{bmatrix},
\quad
N^{b_1}_i N^{b_2}_j = 
\begin{bmatrix}
1  &0 &0 &0\\
0  &1 &b_1i +b_2 j &0\\
0 &0 &1 &0\\
0 &0 &0 &1\\
\end{bmatrix}.
\] 

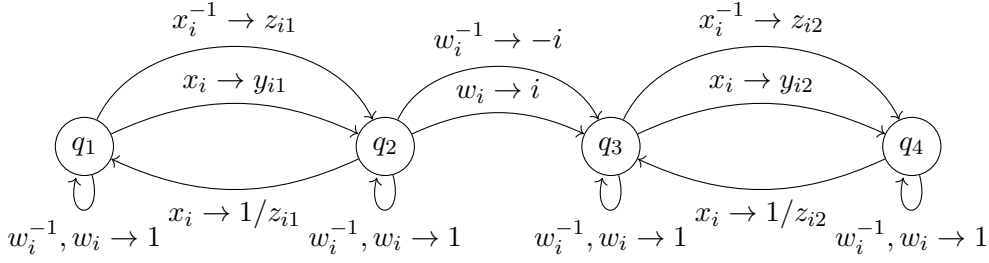
\begin{figure}[h]
\begin{center}
\begin{tikzpicture}
\node(pseudo) at (-1,0){};
\node(1) at (-3.5,0)[shape=circle,draw]        {$q_1$};
\node(2) at (.5,0)[shape=circle,draw]        {$q_2$};
\node(3) at (3.5,0)[shape=circle,draw] {$q_3$};
\node(4) at (7.5,0)[shape=circle,draw] {$q_4$};
\path [->]


  (2)      edge [bend right=-25]  node [above]  {$w_i\rightarrow i$}     (3)
  (2)      edge [bend right=-65]  node [above]  {$w^{-1}_i\rightarrow -i$}     (3)
  
  
(1)      edge [loop below]    node [below]  {$w^{-1}_i,w_i \rightarrow 1$}     ()
(2)      edge [loop below]    node [below]  {$w^{-1}_i ,w_i\rightarrow 1$}     ()
(3)      edge [loop below]    node [below]  {$w^{-1}_i, w_i \rightarrow 1$}     ()
(4)      edge [loop below]    node [below]  {$w^{-1}_i , w_i\rightarrow 1$}     ()

 (1)      edge [bend right=-25]  node [above]  {$x_i\rightarrow  y_{i1}$}     (2)
  (1)      edge [bend right=-65]  node [above]  {$x^{-1}_i\rightarrow z_{i1}$}     (2)
  (2)      edge [bend left=25]  node [below]  {$x_i \rightarrow 1/z_{i1}$}     (1)
  
  (3)      edge [bend right=-25]  node [above]  {$x_i\rightarrow  y_{i2}$}     (4)
  (3)      edge [bend right=-65]  node [above]  {$x^{-1}_i\rightarrow z_{i2}$}     (4)
  (4)      edge [bend left=25]  node [below]  {$x_i \rightarrow 1/z_{i2}$}     (3);

\end{tikzpicture}
\caption{The transition diagram of the automaton}\label{fig5}
\end{center} 
\end{figure}

Hence, evaluating $f(N_1M_1N_1, N_2M_2N_2)$ we obtain (a polynomial
weakly equivalent to) $\varphi(f)$ at the $(1,4)^{th}$ entry. The
complete automaton is depicted in figure~\ref{fig5}.

We now explain the general construction. For $f\in \F\angle{X,X^{-1}}$
let $H_{\ell}(f)$ denote the degree-$\ell$ homogeneous part of $f$. We
will denote by $\widehat{\varphi(H_{\ell}(f))}$ an arbitrary
polynomial in $\F[Y,Z]$ weakly equivalent to $\varphi(H_{\ell}(f))$.

\begin{lemma}\label{encoding}
Let $f\in\F\angle{X,X^{-1}}$ be a nonzero expression of degree
$d$. There is an $n$-tuple of $2d\times 2d$ matrices $(M_1,
M_2,\ldots,M_n)$ whose entries are either scalars, or variables
$u\in Y\cup Z$, or their inverses $1/u$,  such that
\[
\left(f(M_1,\ldots,M_n)\right)_{1,2d} = \widehat{\varphi(H_{d}(f))}.
\]
Furthermore, for each degree-$d$ reduced word of  $m =
x^{b_1}_{i_1}x^{b_2}_{i_2}\cdots x^{b_d}_{i_d}$ in $\F\angle{X,
  X^{-1}}$,
\begin{equation}\label{scalar-value}
[\varphi(m)]\widehat{\varphi(H_{d}(f))} = [m]f \cdot \prod_{j=1}^{d-1} (b_j\cdot i_j + b_{j+1}\cdot i_{j+1}).
\end{equation}

\end{lemma} 

 \begin{proof}
Let $e_{ij}$, for $i,j\in [k]$, be the $(i,j)^{th}$ elementary matrix
in $\mathbb{M}_k(\F)$: its $(i,j)^{th}$ entry is $1$ and other entries
are $0$.

We now define the transition matrices of the NFA for variables $\{w_i
: 1\leq i \leq n\}$ and $\{x_i:1\leq i\leq n\}$.  For each $i\in [n]$,
define $2\times 2$ matrix $N'_{i} = e_{11} + e_{22} + i\cdot
e_{12}$.  Now $N_i$ is a $2d\times 2d$ matrix defined as the block
diagonal matrix,
 \[
{ N'_{i} = 
 \begin{bmatrix}
 1   &i \\
 0   &1  \\
 \end{bmatrix}
 },
 \quad\quad
{N_{i}=  \begin{bmatrix}
 1 &0 &0 &\ldots &0 &0\\
 0 &N'_{i} &0 &\ldots &0 &0\\
 0 &0 &N'_{i}  &\ldots &0 &0\\
 \vdots &\vdots &\vdots &\ddots  &\vdots &\vdots\\
 0 &0  &0 &\ldots &N'_{i} &0\\
 0 &0  &0 &\ldots &0 &1\\

 \end{bmatrix}
 }.
 \]
 \[
{ N'^{-1}_{i} = 
 \begin{bmatrix}
 1   &-i \\
 0   &1  \\
 \end{bmatrix}
 },
 \quad\quad
{N^{-1}_{i}=  \begin{bmatrix}
 1 &0 &0 &\ldots &0 &0\\
 0 &N'^{-1}_{i}  &0 &\ldots &0 &0\\
 0 &0 &N'^{-1}_{i}   &\ldots &0 &0\\
 \vdots &\vdots &\vdots &\ddots  &\vdots &\vdots\\
 0 &0  &0 &\ldots &N'^{-1}_{i}  &0\\
 0 &0  &0 &\ldots &0 &1\\

 \end{bmatrix}
 }.
 \]

Each $M_i, 1\le i\le n$ is the $2d\times 2d$ block diagonal matrix
where each $2\times 2$ block $M'_{ij}, 1\leq j \leq d$ is a $2\times 2$ matrix defined
as $M'_{i,j} = y_{ij}\cdot e_{12} + \frac{1}{z_{ij}}\cdot
e_{21}$. Their inverses have a similar structure.
 \[
{ M'_{i,p} = 
 \begin{bmatrix}
 0 &y_{ip} \\
 \frac{1}{z_{ip}} &0 \\
 \end{bmatrix}
 },
 \quad\quad
{M_{i}=  \begin{bmatrix}
 M'_{i,1} &0 &0 &\ldots &0 \\
 0  &M'_{i,2} &0 &\ldots  &0\\
 0 &0 &M'_{i,3} &\ldots  &0\\
 \vdots &\vdots &\vdots &\ddots &\vdots\\
 0 &0  &0 &\ldots  &M'_{i,d}\\
 \end{bmatrix}
 }.\]
 
  \[
{ M'^{-1}_{i,p} = 
 \begin{bmatrix}
 0 & z_{ip} \\
 \frac{1}{y_{ip}} &0 \\
 \end{bmatrix}
 },
 \quad\quad
{M^{-1}_{i}=  \begin{bmatrix}
 M'^{-1}_{i,1} &0 &0 &\ldots &0 \\
 0  &M'^{-1}_{i,2} &0 &\ldots  &0\\
 0 &0 &M'^{-1}_{i,3} &\ldots  &0\\
 \vdots &\vdots &\vdots &\ddots &\vdots\\
 0 &0  &0 &\ldots  &M'^{-1}_{i,d}\\
 \end{bmatrix}
 }.
\] 
The corresponding NFA is depicted in Figure~\ref{fig6}. We substitute
each $x_{i_j}$ by the $2d\times 2d$ matrix
$N_{i_j}M_{i_j}N_{i_j}$. Each $x^{-1}_{i_j}$ is substituted by its
inverse matrix $N^{-1}_{i_j}M^{-1}_{i_j}N^{-1}_{i_j}$.

\subsubsection*{Correctness.}  Consider a degree-$d$ reduced word
$m=x^{b_1}_{i_1}x^{b_2}_{i_2}\cdots x^{b_d}_{i_d}$.

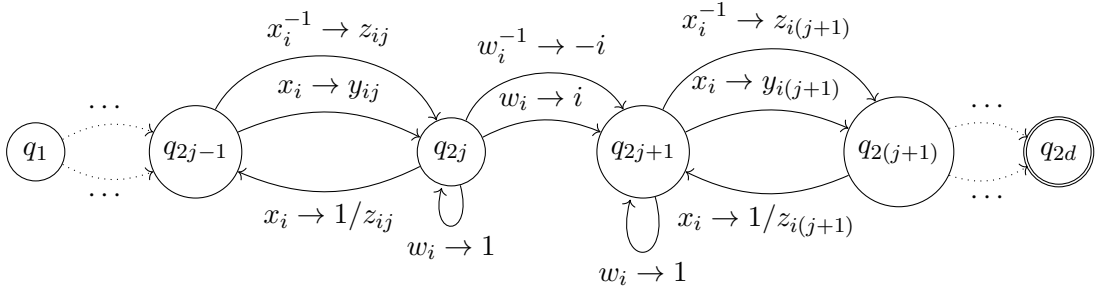
\begin{figure}
\begin{center}
\begin{tikzpicture}[scale=0.85]
\node(pseudo) at (-1,0){}; \node(0) at (-6,0)[shape=circle,draw]
     {$q_1$}; \node(1) at (-3.5,0)[shape=circle,draw] {$q_{2j-1}$};
     \node(2) at (.5,0)[shape=circle,draw] {$q_{2j}$}; \node(3) at
     (3.5,0)[shape=circle,draw] {$q_{2j+1}$}; \node(4) at
     (7.5,0)[shape=circle,draw] {$q_{2(j+1)}$}; \node(5) at
     (10,0)[shape=circle,draw,double] {$q_{2d}$}; \path [->]


  (0)      edge [bend right=-25,dotted]  node [above]  {$\cdots$}     (1)
  (0)      edge [bend right=25,dotted]  node [below]  {$\cdots$}     (1)
  
  (1)      edge [bend right=-25]  node [above]  {$x_i\rightarrow  y_{ij}$}     (2)
  (1)      edge [bend right=-65]  node [above]  {$x^{-1}_i\rightarrow z_{ij}$}     (2)
  (2)      edge [bend left=25]  node [below]  {$x_i\rightarrow  1/z_{ij}$}     (1)
  
  (2)      edge [bend right=-25]  node [above]  {$w_i \rightarrow i$}     (3)
  (2)      edge [bend right=-65]  node [above]  {$w^{-1}_i \rightarrow -i$}     (3)
  
  (3)      edge [bend right=-25]  node [above]  {$x_i\rightarrow y_{i(j+1)}$}     (4)
  (3)      edge [bend right=-65]  node [above]  {$x^{-1}_i\rightarrow  z_{i(j+1)}$}     (4)
  (4)      edge [bend left=25]  node [below]  {$x_i\rightarrow  1/z_{i(j+1)}$}     (3)
  
 (4)      edge[bend right= -25, dotted]                 node [above]  {$\cdots$}     (5)
 (4)      edge[bend right= 25, dotted]                 node [below]  {$\cdots$}     (5)

(2)      edge [loop below]    node [below]  {$w_i \rightarrow 1$}     ()
(3)      edge [loop below]    node [below]  {$w_i \rightarrow 1$}     ();

\end{tikzpicture}
\caption{The transition diagram of the automaton}\label{fig6}
\end{center} 
\end{figure}

Following the automaton construction of Figure~\ref{fig6}, $x^{b_i}_i$
occurring at position $j$ is substituted by $([\mathbbm{1}_{[b_i
      =1]}y_{ij}+ \mathbbm{1}_{[b_i =-1]} z_{ij})$. Moreover, for
  each position $j\in [d - 1]$, the adjacent pair $x^{b_j}_{i_j}
  x^{b_{j+1}}_{i_{j+1}}$ produces a scalar factor $(b_j\cdot i_j +
  b_{j+1}\cdot i_{j+1})$ due to the product
  $N^{b_j}_{i_j}N^{b_{j+1}}_{i_{j+1}}$. Consequently, it follows that
\[
\left( m(M_1,\ldots,M_n)\right)_{1,2d} = \prod_{j=1}^{d-1} (b_j\cdot i_j + b_{j+1}\cdot i_{j+1}) \prod_{j=1}^d ([b_j = 1]y_{i_jj} + [b_j = -1]z_{i_jj} ).   
\]

As $\varphi$ is a linear map, the lemma follows. 
\end{proof}

\subsection{Black-box identity testing for circuits in free group algebra}\label{fga-circ-pit}

Theorem \ref{PI-theorem} follows easily from Lemma~\ref{encoding}.
Lemma \ref{encoding} says that if $f\in \F\angle{\gX}$ is nonzero of
degree $d$ then the $(1,2d)$ entry of the matrix $p(N_1 M_1 N_1,
\ldots, N_n M_n N_n)$ is a nonzero polynomial in $\F[Y,Z]$. 
Hence $f$ can not be an identity for $M_{2d}(\F)$.

\vspace{0.2cm} It also immediately gives an identity testing
algorithm. 
We can randomly
substitute for the variables 
and apply
the Schwartz-Zippel-Demillo-Lipton Theorem~\cite{Sch80, Zip79,
  DL78}. This completes the proof of the Corollary \ref{rand-nccirc}.

\subsection{Reconstruction of sparse expressions in free group algebra}\label{fga-sparse-pit}

If the black-box contains an $s$-sparse expression in $\F\angle{\gX}$,
we give a $\poly(s,n,d)$ deterministic interpolation algorithm (which
also gives a deterministic identity testing for such expressions). We use a result
of Klivans-Spielman~\cite[Theorem11]{KS01} that constructs a test set
in deterministic polynomial time for sparse commutative polynomials,
which is used for the interpolation algorithm.
 
\subsubsection*{Proof of Theorem \ref{det-sparsenc}}

Let the black-box expression $f$ be $s$-sparse of degree $d$. By
Lemma~\ref{encoding}, a polynomial $\widehat{\varphi(H_d (p))}$ in
$\F[Y,Z]$ is obtained at the $(1,2d)^{th}$ entry of the matrix
$f(M_1,\ldots,M_n)$, where $M_i\in\mathbb{M}_{2d}(\F[Y,Z])$ is as
defined in Lemma~\ref{encoding}. By Definition~\ref{defn-encoding},
$\varphi(f)\in \F[Y,Z]$ is $s$-sparse and has $2nd$ variables.  Let
$\mathcal{H}_{2nd,d,s}$ be the corresponding test set from~\cite{KS01}
to interpolate a polynomial of degree $d$ and $s$-sparse over $2nd$
variables. Querying the black-box on
$M_1(\vec{h}),M_2(\vec{h}),\ldots,M_n(\vec{h})$ for each $\vec{h} \in
\mathcal{H}_{2nd,d,s}$ we can interpolate the commutative polynomial
$\widehat{\varphi(H_d(f))}$ and obtain an expression for
$\widehat{\varphi(H_d(f))} = \sum^s_{t=1} c_{m_t} m_t $ as a sum of
monomials.

We now need to adjust the extra scalar factors in
$\widehat{\varphi(H_d(f))}$ to obtain $\varphi(H_d(f))$. We can
perform this adjustment for each monomial  as 
Lemma~\ref{encoding} shows that the extra scalar factor for the word
$m=x^{b_1}_{i_1}x^{b_2}_{i_2}\cdots x^{b_\ell}_{i_\ell}$ is just
$\alpha_m=\prod_{j=1}^{\ell-1} (b_j\cdot i_j + b_{j+1}\cdot
i_{j+1})$. So the algorithm constructs the expression
$\widehat{\varphi(H_d (f))} =\sum^{s}_{t=1}
\frac{c_{m_t}}{\alpha_{m_t}} m_t$. We can remove the factors
$\alpha_{m_t}$ for each monomial $m_t$ and \emph{invert} the map 
$\varphi$ (using the $4^{th}$ property of Definition~\ref{defn-encoding}) on every monomial $m_t$ to obtain $H_{d}(f)$ as a sum of
degree $d$ reduced words. This yields the expression for highest degree homogeneous
component of $f$. We can repeat the above procedure on $f - H_d(f)$
and reconstruct the remaining homogeneous components of $f$. \qed

 \section{Black-box Identity Testing for Expressions of Exponential Degree and Exponential Sparsity}\label{fga-expodeg-pit}
 
 In this section, we prove a different generalization of
 Amitsur-Levitzki theorem~\cite{AL50} for free group algebras, based
 on ideas from \cite{AJMR17}. We show that the dimension of the matrix
 algebra for which a nonzero input expression $f$ does not vanish is
 logarithmic in the sparsity of $f$. It yields a randomized
 $\poly(\log D,\log s, n)$ time identity testing algorithm when the black-box
 contains an expression of degree $D$ and sparsity $s$.
 
We first recall the notion of \emph{isolating index set} from
\cite{AJMR17}.

\begin{definition}
  Let $\mathcal{M} \subseteq \{X,X^{-1}\}^D$ be a subset of reduced
  words of degree $D$. An index set $I\subseteq [D]$ is an
  \emph{isolating index set} for $\mathcal{M}$ if there is a word
  $m\in\mathcal{M}$ such that for each
  $m'\in\mathcal{M}\setminus\{m\}$ there is an index $i\in I$ for
  which $m[i]\ne m'[i]$. I.e.\ no other word in $\mathcal{M}$ agrees
  with $m$ on all positions in the index set $I$. We say $m$ is an
  \emph{isolated word}.
\end{definition}

In the following lemma we show that $\mathcal{M}$ has an isolating
index set of size $\log |\mathcal{M}|$. The proof is identical to
\cite{AJMR17}. Nevertheless, we give the simple details for
completeness because we deal with both variables and their inverses.

\begin{lemma}{\rm\cite{AJMR17}}\label{isolating-lemma}
  Let $\mathcal{M} \subseteq \{X,X^{-1}\}^D$ be reduced degree-$D$
  words. Then $\mathcal{M}$ has an isolating index set of size $k$
  which is bounded by $\log |\mathcal{M}|$.
\end{lemma}

\begin{proof}
  The words $m\in \mathcal{M}$ are indexed, where $m[i]$ denotes the
  variable (or the inverse of a variable) in the $i^{th}$ position of
  $m$. Let $i_1\le D$ be the first index such that not all words agree
  on the $i_1^{th}$ position. Let
\begin{align*}
S^{+}_j & =  \{ m :  m[i_1] = x_j\}\\ 
S^{-}_j & =  \{ m :  m[i_1] = x^{-1}_j\}.  
\end{align*}

For some $j$, $|S^{+}_j|$ or $|S^{-}_j|$ is of size at most
$|\mathcal{M}|/2$.  Let $S^b_{i_1}$ denote that subset,
$b\in\{+,-\}$.  We replace $\mathcal{M}$ by $S^b_{i_1}$ and repeat the
same argument for at most $\log |\mathcal{M}|$ steps. Clearly, by
this process, we identify a set of indices $I=\{i_1, \ldots, i_k'\}$,
$k'\le \log |\mathcal{M}|$ such that the set shrinks to a
singleton set $\{m\}$. Clearly, $I$ is an isolating index set as
witnessed by the \emph{isolating word} $m$.
\end{proof}

\subsection*{Proof of Theorem \ref{sparse-PI-theorem}}
 
Let $k = 4(k' + 1)$ where $k'$ is the size of the isolating set $I$.
As in Section~\ref{main-sect}, we substitute each $x_i$ by
$w_ix_iw_i$, where $w_i, i\in[n]$ are $n$ new variables. The
transition matrices for $w_i$ and $x_i$ are denoted by $N_i$ and $M_i$
respectively.
  
For $1\leq i \leq n$, we define $k\times k$ matrix $N_i$ as a block
diagonal matrix of $k$ many $4\times 4$ matrices $N'_i$ where $N'_i =
I_4 + i(e_{12}+e_{34}+e_{32}+ e_{14})$.
 \[
{ N'_{i} = 
 \begin{bmatrix}
 1   &i  &0 &i\\
 0  &1   &0 &0\\
 0  &i   &1 &i\\
 0 &0 &0 &1
 \end{bmatrix}
 },
 \quad\quad
{N_{i}=  \begin{bmatrix}
 N'_{i} &0 &0 &\ldots &0 \\
 0 &N'_{i} &0 &\ldots &0 \\
 0 &0 &N'_{i}  &\ldots &0 \\
 \vdots &\vdots &\vdots &\ddots  &\vdots\\
 0 &0  &0 &\ldots &N'_{i} \\

 \end{bmatrix}
 },\]

 \[
{ N'^{-1}_{i} = 
 \begin{bmatrix}
 1   &-i  &0 &-i\\
 0  &1   &0 &0\\
 0  &-i   &1 &-i\\
 0 &0 &0 &1
 \end{bmatrix}
 },
 \quad\quad
{N_i^{-1}=  \begin{bmatrix}
 N'^{-1}_{i} &0 &0 &\ldots &0 \\
 0 &N'^{-1}_{i} &0 &\ldots &0 \\
 0 &0 &N'^{-1}_{i}  &\ldots &0 \\
 \vdots &\vdots &\vdots &\ddots  &\vdots\\
 0 &0  &0 &\ldots &N'^{-1}_{i} \\
 \end{bmatrix}
 }.\]

Notice that

\[
{N'^{b_1}_{i}N'^{b_2}_{j}=  
\begin{bmatrix}

 1   &(b_1 i + b_2 j)  &0 &(b_1 i + b_2 j)\\

 0  &1   &0 &0\\

 0  &(b_1 i + b_2 j)   &1 &(b_1 i + b_2 j)\\

 0 &0 &0 &1
\end{bmatrix}
}.
\]
 
We now define the $k\times k$ transition matrix $M_i$ as a block
diagonal matrix,
 
  \[
{ M'_{i,j} = 
 \begin{bmatrix}
 0 &y_{ij} \\
 \frac{1}{z_{ij}} &0 \\
 \end{bmatrix}
 },\quad\quad
 { M'_{\xi_i} = 
 \begin{bmatrix}
 0 &\xi_{i} \\
 \frac{1}{\xi_{i}} &0 \\
 \end{bmatrix}
 },\]
 \[
{M_{i}=  \begin{bmatrix}
 1 &0 &0 &0 &\ldots &0 &0 \\
 0 &M_{\xi_1} &0 &0 &\ldots  &0 &0\\
 0 &0 &M'_{i,1} &0  &\ldots &0  &0 \\
 0 &0 &0 &M_{\xi_2}  &\ldots  &0 &0 \\
 \vdots &\vdots &\vdots &\vdots &\ddots  &\vdots &\vdots\\
0 &0  &0  &0 &\ldots &M_{\xi_{k'+1}} &0 \\
 0 &0  &0  &0 &\ldots &0 &1 \\
 \end{bmatrix}.
 }\]
 
These matrices can be seen as the transitions of a suitable NFA. We
sketch the construction of this NFA.
 
\begin{figure}[h]
\begin{center}
\begin{tikzpicture}[scale=0.8]
\node(pseudo) at (-1,0){};
\node(1) at (-17,0)[shape=circle,draw] {$ \text{Start}$};

\node(2) at (-14.5,0)[shape=circle,draw] {$ \text{Skip 1}$};
\node(3) at (-12,0)[shape=circle,draw] {$ \text{Enc 1}$};
\node(4) at (-9.5,0)[shape=circle,draw] {$ \text{Skip 2}$};
\node(5) at (-7,0)[shape=circle,draw] {$ \text{Enc 2}$};
\node(6) at (-3.5,0)[shape=circle,draw] {$ \text{Skip $k'$}$};
\node(7) at (-1,0)[shape=circle,draw] {$ \text{Enc $k'$}$};
\node(8) at (1,0)[shape=circle,draw,double] {$ \text{Final}$};

\path [->]
  
  (1)      edge [bend right=-25]  node [above]  {} (2)
  (1)      edge [bend right=-65]  node [above]  {} (3)
  (2)      edge [bend left=25]  node [below]  {} (3)
  (3)      edge [bend left=25]  node [below]  {} (4)
  (4)      edge [bend left=25]  node [below]  {} (5)
  (3)      edge [bend right=-65]  node [above]  {} (5)
  (5)      edge [dotted]  node [above]  {}     (6)
  (5)      edge [dotted,bend left=75]  node [above]  {}     (6)
  (6)      edge [bend right=-25]  node [above]  {} (7)
  (6)      edge [bend right=-65]  node [above]  {} (8)
  (7)      edge [bend left=25]  node [below]  {} (8);

\end{tikzpicture}
\caption{The transition diagram of the automaton}\label{fig9}
\end{center} 
\end{figure}
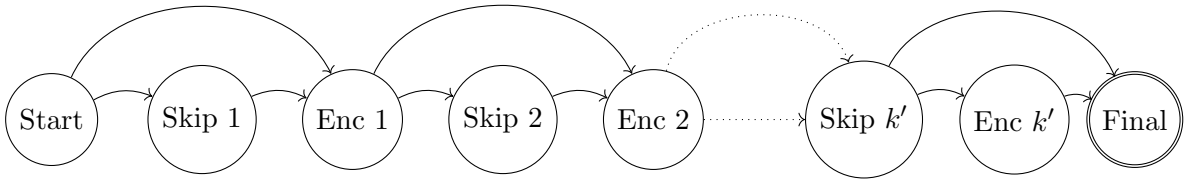

Let $I = \{i_1,\ldots,i_{k'}\}$ be an isolating set such that $i_1<
\ldots< i_{k'}$. Intuitively, the NFA does one of two operations on
each symbol (a variable or its inverse) of the input expression: a
\emph{Skip} or an \emph{Encode}. In a \emph{Skip} stage, the NFA deals
with positions that are not part of the (guessed) isolating index
set. In this stage, the NFA substitutes the $w_i$ variables by
suitable scalars (coming from the $N'_i$ matrices) and $x_i$ variables
by block variables $\{\xi_1,\ldots\xi_{k'+1}\}$. The NFA
nondeterministically decides whether the \emph{Skip} stage is over and
it enters the \emph{Encode} stage for a guessed index of the isolating
set. It substitutes $x_i$ and $x^{-1}_i$ variables by $y_{ij}$ and
$z_{ij}$ respectively. Fig.~\ref{fig9} summarizes the action of the
NFA.

Define $\hat{f}$ in $\F(Y,Z,\overline{\xi})$ to be rational function we obtain at
the $(1,k)^{th}$\footnote{Recall that $k=4(k'+1)$ where $k'$ is the size of an isolating set.} entry by evaluating the expression
$f(N_1M_1N_1,\ldots,N_nM_nN_n)$. Notice that, the isolating word
$m$ of degree $D$ will be of following form $m = W_1 x^{b_{i_1}}_{i_1}
W_2 x^{b_{i_2}}_{i_2} \cdots W_k' x^{b_{i_k'}}_{i_k'} W_{k'+1}$ where
each subword $W_j = x^{b_1}_{j_1}x^{b_2}_{j_2}\cdots
x^{b_{\ell_j}}_{j_{\ell_j}}$ is of length $\ell_j\ge 0$, where some of
the $W_j$ could be the empty word as well. 

We refer to an NFA transition $q_i\to q_j$ as a \emph{forward edge} if
$i<j$ and a \emph{backward edge} if $i>j$. We classify the backward edges in three categories based on the substitution on the edge-label. We say, a backward edge is of \emph{type A} if a variable is substituted by a scalar value; 
a backward edge is of \emph{type B} if a variable is substituted by $\frac{1}{\xi_j}$ for some $j$;
a backward edge is of \emph{type C} if a variable is substituted by $\frac{1}{y_{ij}}$ or $\frac{1}{z_{ij}}$ for some $i,j$.

Consider a walk of the NFA
on an input word $m$ that reaches state $k$ using only \emph{type A}
backward edges. In that case, $m$ is substituted by $\alpha\cdot \hat{m}$ where $\hat{m}$ is a monomial over $\{Y,Z,\xi\}$ of same degree,
\[
\hat{m} = \prod_{j=1}^{k'+1} \xi^{\ell_j}_j \cdot 
\prod_{j=1}^{k'} ([b_{i_j} = 1] y_{i_jj} + [b_{i_j} = -1] z_{i_jj}).
\]
and $\alpha$ is some nonzero constant obtained as a product of $[m]f$
with the scalars obtained as substitutions from the edges
involving the $w_i$ variables in the \emph{Skip} stages. Indeed, as we
can see from the entries of product matrices $N'^{b_1}_i\cdot N'^{b_2}_j$,
where $b_1,b_2\in\{-1,1\}$, the scalar $\alpha$ is a product of $[m]f$ with
terms of the form $b_1i+b_2j$, for $i\ne j$, each of which is nonzero for any reduced word.

\begin{figure}[h]
\begin{center}
\begin{tikzpicture}
\node(pseudo) at (-1,0){};
\node(0) at (0,0)[shape=circle,draw]        {$q_{4j}$};
\node(1) at (3,0)[shape=circle,draw]        {$q_{4j+1}$};
\path [->]  
 (0)      edge [bend right=-25]  node [above]  {$x_i, x^{-1}_i\rightarrow y_{ij}, z_{ij}$}     (1)
  (1)      edge [bend left=25]  node [below]  {$x_i, x^{-1}_i\rightarrow 1/z_{ij}, 1/y_{ij}$}     (0);
\end{tikzpicture}
\caption{The transition diagram of the automaton at \emph{Encode} stage}\label{fig7}
\end{center} 
\end{figure}
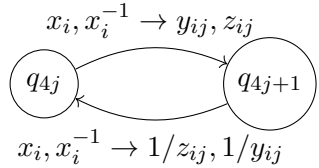

\begin{figure}[h]
\begin{center}
\begin{tikzpicture}
\node(pseudo) at (-1,0){};
\node(1) at (-3.5,0)[shape=circle,draw]        {$q_{4j-3}$};
\node(2) at (.5,0)[shape=circle,draw]        {$q_{4j-2}$};
\node(3) at (3.5,0)[shape=circle,draw] {$q_{4j-1}$};
\node(4) at (7.5,0)[shape=circle,draw] {$q_{4j}$};
\path [->]

  (1)      edge [bend right=-25]  node [below]  {$w_i\rightarrow i$}     (2)
  (1)      edge [bend right=-45,font=\fontsize{9.5}{1}]  node [above]  {$w^{-1}_i\rightarrow -i$}     (2)
  (1)      edge [bend right=-65]  node [above]  {$w_i\rightarrow i, w^{-1}_i\rightarrow -i$}     (4)
  (2)      edge [bend right=-25]  node [above]  {$x_i\rightarrow \xi_{j}$}     (3)
  (3)      edge [bend left=75]  node [below]  {$x_i\rightarrow \frac{1}{\xi_j}$}     (2)
  (3)      edge [bend left=20]  node [above]  {$w_i\rightarrow i$}     (2)
  (3)      edge [bend left=55 ,font=\fontsize{9.5}{1}]  node [above]  {$w^{-1}_i\rightarrow -i$}     (2)
  (3)      edge [bend right=-25 ]  node [below]  {$w_i\rightarrow i$}     (4)
  (3)      edge [bend right=-45 ,font=\fontsize{9.5}{1}]  node [above]  {$w^{-1}_i\rightarrow -i$}     (4)
  
 (1)      edge [loop below]    node [below]  {$w_i \rightarrow 1$}     () 
(2)      edge [loop above]    node [above]  {$w_i \rightarrow 1$}     ()
(3)      edge [loop above]    node [above]  {$w_i \rightarrow 1$}     ()
(4)      edge [loop below]    node [below]  {$w_i \rightarrow 1$}     ();

\end{tikzpicture}
\caption{The transition diagram of the automaton at \emph{Skip} stage}\label{fig8}
\end{center} 
\end{figure}
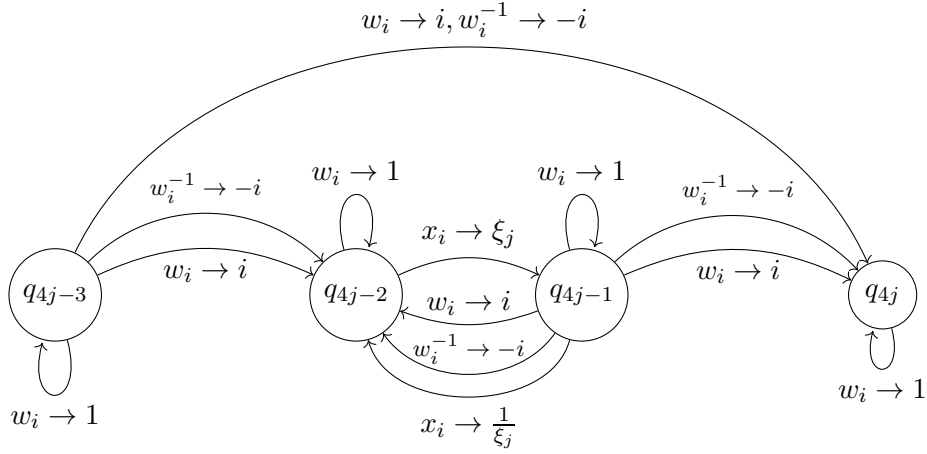

\begin{claim}
\[
[\hat{m}]\hat{f} \ne 0 \textrm{ iff } [m]f\ne 0.
\]
\end{claim} 

 \begin{proof}
 It suffices to show that for any word $m'\ne m$, where $m'$ has
 degree $\leq D$, no walks of the NFA accepting $m'$ generate
 $\hat{m}$ after substitution. We now argue that no other walks in the
 NFA can generate $\hat{m}$.  For a computation path $J$, the monomial
 $m_J$ in $\hat{f}$ has two parts, let us call it $skip_J$ and
 $encode_J$ where $skip_j$ is a monomial over
 $\{\xi_1,\ldots,\xi_{k'+1}\}$ and $encode_j$ is a monomial over
 $\{y_{i,j},z_{i,j}\}_{i\in[n], j\in [k']}$. If the computation path
 $J$ (which is different from the computation path described above for
 $\hat{m}$) uses only \emph{type A} backward edges, then necessarily $m_J\neq
 \hat{m}$ from the definition of \emph{isolating index set}. This
 argument is analogous to the argument given in \cite{AJMR17}.
 
Now consider a walk $J$ which involves backward edges of other types. Let us first
consider those walks that take backward edges only of \emph{type A} and \emph{type B}.
Such a walk still produces a monomial over
$\{y_{i,j},z_{i,j}\}_{i\in [n], j\in [k']}$ and $\{\xi_i\}_{1\leq i
  \leq k'+1}$ because division only by $\xi_i$ variables occur in the
resulting expression. Since $\hat{m}$ is of highest degree, the total
degree of these monomials is strictly lesser than degree of
$\hat{m}$. For those walks that take at least one backward edge of \emph{type C}, 
a rational expression in
$\{y_{i,j},z_{i,j}\}_{i\in [n], j\in [k']}$ and $\{\xi_i\}_{1\leq i
  \leq k'+1}$ is produced (as there is division by $y_{ij}$ or
$z_{ij}$ variables). As the sum of the degree of the numerator and
degree of the numerator is bounded by the total degree, the degree of
the numerator is smaller than degree of $\hat{m}$.

Thus the $(1,k)^{th}$ entry of the output matrix is of the form
$\sum_{i=1}^{N_1} c_im_i + \sum_{j=1}^{N_2} r_j$ where
$\{m_1,\ldots,m_{N_1}\}$ are monomials arising from different walks
(w.l.o.g. assume that $m_1 = \hat{m}$) and $\{r_1,\ldots, r_{N_2}\}$
are the rational expressions from the other walks (due to the backward
edges of \emph{type C}). 
Note that, denominator in each $r_j$ is a monomial over $Y,Z$ of degree at most $D$. Let $L = \prod_{i=1}^n \prod_{j=1}^{k'} y^D_{i,j} \cdot  z^D_{i,j}$.
Now, we have,
\[
\sum_{i=1}^{N_1} c_im_i + \sum_{j=1}^{N_2} r_j = \frac{1}{L}\cdot
\left(\sum_{i=1}^{N_1} c_im_iL + \sum_{j=1}^{N_2} p_j\right).
\]
Since $\hat{m}L \neq m_i L$ for any $i\in \{2,\ldots,N_1\}$ and degree
of each $p_j<$ degree of $\hat{m}L$ for any $j\in \{1,\ldots,N_2\}$,
the numerator of the final expression is a nonzero polynomial in
$\F[Y,Z,\overline{\xi}]$.
\end{proof}

The above proof shows that the matrix $f(N_1M_1N_1,\ldots,N_nM_nN_n)$
is nonzero with rational entries in $\F[Y,Z,\overline{\xi}]$. Each
entry is a linear combination of terms of the form $m_1/m_2$, where
$m_1$ and $m_2$ are monomials in $Y\cup Z\cup\{\xi_1,\ldots,\xi_{k'+1}\}$
of degree bounded by $D$. This completes the proof. \qed 

To get an identity testing algorithm, we can do random substitutions.
The matrix dimension is $\log s$ and the
overall running time of the algorithm is $\poly(n,\log s,\log D)$.
This also proves Corollary \ref{expo-deg-sparse-pit-theorem}.
\qed

\begin{remark}\label{useful}
For algorithmic purposes, we note that Theorem~\ref{PI-theorem} is
sometimes preferable to Theorem \ref{sparse-PI-theorem}. For instance,
the encoding used in Theorem \ref{sparse-PI-theorem} does not preserve
the sparsity of the polynomial as required in the sparse
reconstruction result (Theorem~\ref{det-sparsenc}).
\end{remark}

\section{Adaptation for Fields of Positive Characteristic}\label{finite-field-modifications}

Let $\F$ be any finite field of characteristic $p$. We need to ensure that for each word
$m$ in the free group algebra, the scalar $\alpha_m$ (see Equation~\ref{scalar-value}) produced by the
automaton  described in Section~\ref{main-sect}
is not zero in $\F$.  Recall that, reading $w^{b_i}_{i}
w^{b_{j}}_{j}$ for two consecutive positions, the automaton
produces a scalar $(b_i\cdot i + b_{j}\cdot j)$ where $b_i,b_j\in
\{-1,+1\}$. Moreover, this is the only way the automaton produces a
scalar and for each $m$, $\alpha_m$ is a product of such terms.  Hence, all we need to ensure is that for each pair $i,j\in
[n]$, $(b_i\cdot i + b_{j}\cdot j)\neq 0$. Similarly, it ensures that the scalar produced by the automaton described in Section~\ref{fga-expodeg-pit} is non-zero.

We note that, if $p$ is more than $2n$ then each term $(b_i\cdot i + b_j\cdot j) \neq 0 \pmod p$ where $b_i,b_j \in \{ -1 , +1\}$ and $i,j \in [n]$. 
This results in a dependence on the characteristic of the base field for the analogous statements of Theorems~\ref{PI-theorem},~\ref{sparse-PI-theorem} over finite field.
Additionally, for Theorem \ref{PI-theorem}, the $(1,2d)^{th}$ entry of the output matrix is a polynomial of degree $d$, and for Theorem \ref{sparse-PI-theorem}, 
the degrees of the numerator polynomials in the rational expression of the output matrix is bounded by some scalar multiple of $nD \log s$. 
This lower bounds the size of the fields in the 
application. We summarize the above discussion in the following. 

\begin{obs}\label{finite-field-1}
We can obtain results analogous to Theorem~\ref{PI-theorem} and Theorem~\ref{sparse-PI-theorem} over finite fields of characteristic more than $2n$ 
and sizes at least $d+1$ or $\Omega(nD\log s)$ respectively. 
\end{obs}    

However, the algorithms presented in Theorem~\ref{det-sparsenc} and
Corollaries~\ref{rand-nccirc},~\ref{expo-deg-sparse-pit-theorem} can
be modified to work for finite fields of any characteristic. To this
end, we first notice the following simple fact.

\begin{proposition}\label{onepisigma}
Let $\F$ be a finite field of characteristic $p\leq 2n$. In We can find
elements $\alpha_1,\alpha_2,\ldots,\alpha_n$ from a suitable
(deterministically constructed) small extension field $\F'$ of $\F$ in
deterministic $\poly(n)$ time, such that for any $b_i\in\{-1,1\}, 1\le
i\le n$ we have
\[
 \text{For each } \ 1\leq i<j \leq n, 
\ b_i\alpha_i + b_j\alpha_j \neq 0.
\]
\end{proposition}

Let $\alpha_1,\alpha_2,\ldots,\alpha_n\in\F'$ as given by the above
proposition. We modify the matrix ${N'}_i$ in the proof of
Theorem~\ref{det-sparsenc} and Corollary~\ref{rand-nccirc} as
\[
 { N'_{i} = 
 \begin{bmatrix}
 1   &\alpha_i \\
 0   &1  \\
 \end{bmatrix}
 },
\]
and in Corollary~\ref{expo-deg-sparse-pit-theorem} we modify ${N'}_i$ as 
\[
{ N'_{i} = 
 \begin{bmatrix}
 1   &\alpha_i  &0 &\alpha_i\\
 0  &1   &0 &0\\
 0  &\alpha_i   &1 &\alpha_i\\
 0 &0 &0 &1
 \end{bmatrix}
 }.
 \]
For each pair $i,j\in [n]$, $(b_i\cdot \alpha_i + b_{j}\cdot
\alpha_j)\neq 0$ by Proposition~\ref{onepisigma}. Thus, for each word
$m$, the scalar $\alpha_m$ produced by the automata are nonzero in the
extension field $\F'$ as well. Furthermore, the test set of
\cite{KS01} works for all fields. Hence Theorem \ref{det-sparsenc}
holds for all finite fields too.  To obtain Corollaries
\ref{rand-nccirc} and \ref{expo-deg-sparse-pit-theorem}, we need to do
the random substitution from suitable small degree extension fields
and use Schwartz-Zippel-Demillo-Lipton Theorem~\cite{Sch80, Zip79,
  DL78}. In summary, our algorithms in the paper can be adapted to
work over all fields.

\begin{pproof}
 Define polynomial $g\in \F[x_1,x_2,\ldots,x_n]$ as
 \[
  g(x_1,x_2,\ldots,x_n) = \prod_{1\leq i < j \leq n} (x_i + x_j) \cdot  (x_i - x_j).
 \]

We substitute $y^i$ for $x_i, 1\le i\le n$. Then
$g(y,y^2,\ldots,y^n)=G(y)\in\F[y]$ is a univariate polynomial of
degree at most $2n^3$. Using standard techniques, in deterministic
polynomial time we can construct an extension field $\F'$ of $\F$ such
that $|\F'|$ is of $\poly(n)\ge 2n^3+1$ size. We can find an element
$\alpha\in\F'$ such that $G(\alpha)\ne 0$ and set $\alpha_i=\alpha^i,
1\le i\le n$. \qed
\end{pproof}

 \bibliographystyle{alpha}

\end{document}